\documentclass[12pt]{article}

\usepackage{bbm,latexsym,epsfig}

\newcommand{\be}{\begin{equation}}
\newcommand{\ee}{\end{equation}}
\newcommand{\ba}{\begin{eqnarray}}
\newcommand{\ea}{\end{eqnarray}}
\newcommand{\no}{\nonumber\\}

\newcommand{\lesssim}{ \ \mbox{\raisebox{-3pt}{$\stackrel%
{\displaystyle <}{\sim}$}} \ }

\begin{document}

\title{\normalsize \hfill UWThPh-2006-6 \\[1cm]
\LARGE
Charged-fermion masses in $SO(10)$: \\
analysis with scalars in $\mathbf{10} \oplus \mathbf{120}$}
\author{
Lu\'\i s Lavoura\thanks{E-mail: balio@cftp.ist.utl.pt} \\
\small Universidade T\'ecnica de Lisboa
and Centro de F\'\i sica Te\'orica de Part\'\i culas \\
\small Instituto Superior T\'ecnico,
1049-001 Lisboa, Portugal
\\*[3.6mm]
Helmut K\"uhb\"ock,\thanks{E-mail: helmut.kuehboeck@gmx.at} \
Walter Grimus\thanks{E-mail: walter.grimus@univie.ac.at} \\
\small Institut f\"ur Theoretische Physik, Universit\"at Wien \\
\small Boltzmanngasse 5, A--1090 Wien, Austria
\\*[4.6mm]}

\date{21 June 2006}

\maketitle

\begin{abstract}
We consider the scenario in which
the mass matrices of the charged fermions
in the $SO(10)$ Grand Unified Theory
are generated exclusively
by renormalizable Yukawa couplings 
to one $\mathbf{10} \oplus \mathbf{120}$ representation of scalars.
We analyze,
partly analytically and partly numerically,
this scenario in the three-generations case.
We demonstrate that it leads to unification
of the $b$ and $\tau$ masses at the GUT scale.
Testing this scenario against the mass values at the GUT scale, 
obtained from the renormalization-group evolution 
in the minimal SUSY extension of the Standard Model,
we find that it is not viable: 
either the down-quark mass or the top-quark mass
must be unrealistically low.
If we include the CKM mixing angles in the test,
then,
in order that the mixing angles are well reproduced,
either the top-quark mass or the strange-quark mass
together with the down-quark mass must be very low.
We conclude that,
assuming a SUSY $SO(10)$ scenario,
charged-fermion mass generation
based exclusively on one $\mathbf{10} \oplus \mathbf{120}$
representation of scalars
is in contradiction with experiment.
\end{abstract}

\newpage

\section{Introduction}

All the fermionic multiplets of one Standard-Model generation,
plus one right-handed neutrino singlet,
fit exactly into the 16-dimensional irreducible representation
of the Grand-Unification group $SO(10)$.
This is the unique and distinguishing feature
of the unified gauge theories (GUTs)
based on this group~\cite{fritzsch}. 
As a bonus,
the presence of three
(for three generations,
as we shall assume in this paper)
right-handed neutrino singlets
allows one to incorporate into this GUT
the seesaw mechanism of type~I~\cite{seesaw}.

However,
when it comes to the scalar sector and to fermion mass generation
the uniqueness of the $SO(10)$ GUT is lost
and numerous ramifications exist.
One possible strategy to limit the freedom in the scalar sector
is to confine oneself to renormalizable terms---for a review see,
for instance,
\cite{senjanovic}.
In that case,
the scalar representations coupling to the fermions
are determined by the relation~\cite{sakita,slansky}
\be
\mathbf{16} \otimes \mathbf{16}
= \left( \mathbf{10} \oplus \mathbf{126} \right)_\mathrm{S}
\oplus \mathbf{120}_\mathrm{AS},
\label{tensor}
\ee
where the subscripts ``S'' and ``AS'' denote,
respectively,
the symmetric and antisymmetric parts of the tensor product.
Thus,
scalars with renormalizable Yukawa couplings to the fermions
must transform under $SO(10)$ either as $\mathbf{10}$,
$\overline{\mathbf{126}}$,
or $\mathbf{120}$
(the $\mathbf{10}$ and $\mathbf{120}$ are real representations;
the $\mathbf{126}$ is complex).
A minimal supersymmetric (SUSY) scenario---which has built-in
the gauge-coupling unification
of the minimal SUSY extension of the Standard Model
(MSSM)---making use of one $\mathbf{10}$
and one $\overline{\mathbf{126}}$ 
for the Yukawa couplings~\cite{babu92} 
has recently received a lot of attention.
This attention was triggered by the observation~\cite{bajc02} that
maximal atmospheric neutrino mixing may in this theory
be related to $b$--$\tau$ unification
via the type~II seesaw mechanism~\cite{typeII}.
Detailed and elaborate studies of this minimal theory
have been performed
for its Yukawa couplings~\cite{goh,bertolini,malinsky,macesanu} 
and scalar potential~\cite{aulakh03,aulakh04}.
This ``minimal SUSY $SO(10)$ GUT'' works very well,
since its Yukawa couplings are able to fit
all fermion masses and mixings,
allowing in particular for small quark mixings
simultaneously with large leptonic mixings.
However,
in this context a minimal Higgs scalar sector
is too constrained~\cite{bajc05}
and does not allow to produce large enough neutrino masses~\cite{garg}.

As a way out,
the $\mathbf{120}$ scalar representation---which had been
somewhat arbitrarily left out---may be used
for a rescue~\cite{aulakh06,bajc06}. 
In~\cite{bertolini,malinsky} that representation
was only taken as a perturbation of the minimal scenario,
to cure minor deficiencies in the fermionic sector.
However,
in~\cite{oshimo} it was pointed out that
the antisymmetric coupling matrix of the $\mathbf{120}$
could be responsible for the different features
of quark and lepton mixing,
since that matrix has different weights
in all four Dirac-type mass matrices---i.e.\ in the Dirac mass matrices
for the up-type quarks,
down-type quarks,
charged leptons,
and neutrinos.
Thinking along this line, 
the roles of the $\mathbf{120}$ and $\overline{\mathbf{126}}$ 
could be interchanged in the charged-fermion sector:
the brunt could be borne by one $\mathbf{10} \oplus \mathbf{120}$,
and the Yukawa couplings of the $\overline{\mathbf{126}}$
would be just a perturbation.
This thought is realized in the model of~\cite{aulakh06},
where the scalar $\overline{\mathbf{126}}$
is still a protagonist in the neutrino sector,
through the type~I seesaw mechanism.

In this paper we investigate
the extreme form of this scenario of~\cite{aulakh06}, 
namely we assume that the $\overline{\mathbf{126}}$
plays no role whatsoever in the 
Yukawa couplings to the charged fermions,\footnote{This idea
was previously put forward in~\cite{matsuda}.}
and may be important only in the neutrino sector,
where it would responsible for large Majorana neutrino masses.
Thus,
we base our investigation on the following assumptions:
\begin{enumerate}
\renewcommand{\labelenumi}{\roman{enumi})}
\item The charged-fermion mass matrices
result solely from the Yukawa couplings of one $\mathbf{10}$
and one $\mathbf{120}$ scalar multiplets.
\item The mechanism for the generation of the light-neutrino mass matrix
is the type~I seesaw mechanism,
possibly with some admixture of type~II.
\end{enumerate}
Due to the first assumption,
the Yukawa-coupling matrix of the $\overline{\mathbf{126}}$
can be used freely for the neutrino mass matrix and,
therefore,
one can accommodate any neutrino masses and lepton mixing that one wants,
through either the type~I or type~II seesaw mechanisms. 
The tight connection between the charged-fermion
and neutrino sectors is lost,
and the predictive power of the model for the neutrino sector too.
The subject of this paper is then only
the discussion of the charged-fermion masses and of quark (CKM) mixing
under the assumption i),
and the working out of where this assumption is successful
and where it might fail.

The charged-fermion sector and the Yukawa couplings in~\cite{bajc06}
coincide with ours.
Still,
our results do \emph{not}, in general, apply to that model.
The reason is that its authors assume split supersymmetry,
where the renormalization-group evolution of the fermion masses
differs from the one of the MSSM.
Indeed,
in order to test any specific scenario
one must use the charged-fermion masses
and the quark mixings at the GUT scale.
Having in mind a SUSY $SO(10)$ GUT and the MSSM,
we use in this paper the values computed in~\cite{das} 
with the renormalization-group evolution of the MSSM.

This paper is organized as follows.
In Section~\ref{masses} we discuss the mass matrices
and count the number of parameters.
Basis-invariant quantities
are introduced in Section~\ref{masses+invariants}.
The derivation of some inequalities,
and $b$--$\tau$ unification,
are discussed in Section~\ref{inequalities}. 
In Section~\ref{luis} we show that
a partly analytical treatment of our scenario is possible
when the Yukawa-coupling matrices are assumed to be real.
Section~\ref{fitting} explains our procedure for the numerical fit
of the mass matrices to the fermion masses and to the CKM mixing angles
at the GUT scale.
We present our results in Section~\ref{results},
which is followed by a brief summary in Section~\ref{concl}.

\section{The charged-fermion mass matrices}
\label{masses}

The mass Lagrangian that we are concerned with is
\be
\mathcal{L}_M = 
- \bar d_L M_d d_R - \bar \ell_L M_\ell \ell_R - 
\bar u_L M_u u_R + \mbox{H.c.}
\label{LM}
\ee
The symmetric and antisymmetric Yukawa couplings 
of one $\mathbf{10}$ and one $\mathbf{120}$ scalar representations,
respectively~\cite{sakita},
generate the mass matrices,
which at the GUT scale may be parametrized as
\ba
M_d    & = & S + e^{i\psi} A, \no
M_\ell & = & S + r e^{i\theta} A, \label{m} \\
M_u    & = & p S + q e^{i\xi} A,  \nonumber 
\ea
$S$ being symmetric while $A$ is antisymmetric.
The parameters $p$,
$q$,
and $r$ are real and positive.
The matrix $S$ is proportional to the Yukawa-coupling matrix
of the $\mathbf{10}$,
while $A$ is proportional to the Yukawa couplings of the $\mathbf{120}$.
The factors $e^{i \psi}$,
$r e^{i \theta}$,
$p$,
and $q e^{i \xi}$ depend on some ratios of vacuum expectation values.

We may perform changes of weak basis
\be
\begin{array}{rcl}
S &\to& U S\, U^T,
\\*[1mm]
A &\to& U A\, U^T,
\end{array}
\label{U}
\ee
where $U$ is unitary.
In this way we may reach convenient weak bases.
We may for instance use $U$ to diagonalize $S$:
\be
S = \left( \begin{array}{ccc} a & 0 & 0 \\ 0 & b & 0 \\ 0 & 0 & c
\end{array} \right),
\quad
A = \left( \begin{array}{ccc}
0 & z & -y \\ -z & 0 & x \\ y & -x & 0
\end{array} \right),
\label{basis1}
\ee
with real and non-negative $a$,
$b$,
and $c$.
Alternatively,
we may use $U$ to force $A$ to have only two non-zero matrix elements,
and moreover two matrix elements of $S$ to vanish:
\be
S = \left( \begin{array}{ccc} a & f & 0 \\ f & b & d \\ 0 & d & c
\end{array} \right),
\quad
A = \left( \begin{array}{ccc}
0 & 0 & 0 \\ 0 & 0 & x \\ 0 & -x & 0
\end{array} \right).
\label{basis2}
\ee
In the weak basis~(\ref{basis2}),
again,
we may choose $a$,
$b$,
and $c$ to be real and non-negative.

As for the number of degrees of freedom
in the mass matrices~(\ref{m}),
we consider two cases:
\begin{itemize}
\item In the complex-Yukawa-couplings case,
the matrices $S$ and $A$ are {\it a priori} complex
and contain nine independent matrix elements
(six in $S$ and three in $A$),
hence nine moduli and nine phases.
One of three phases $\psi$,
$\theta$,
and $\xi$ may be absorbed in the definition of $A$.
Through a weak-basis transformation we may eliminate
the three moduli and six phases which parametrize $U$.
In that case the model has,
therefore,
nine real parameters and five phases.
\item In the real-Yukawa-couplings case,
in which CP violation is considered to be spontaneous,
the matrices $S$ and $A$ are {\it a priori} real
and contain nine independent moduli.
If we want to preserve the reality of $S$ and $A$,
the matrix $U$ of the weak-basis transformation~(\ref{U})
must be chosen real (orthogonal),\footnote{If $S$ and $U$
are assumed to be real,
then one may obtain a weak basis of the form~(\ref{basis1}),
but $a$,
$b$,
and $c$ must be allowed to be negative.
\label{thefoot}
It is only when we allow $U$ to include some $i$ factors
that we may obtain non-negative $a$,
$b$,
and $c$;
but then $x$,
$y$,
and $z$ will not necessarily be real.}
hence it contains three real parameters.
One ends up with nine real parameters as before,
but only three phases.
\end{itemize}
With these 14 (in the complex case)
or 12 (in the real case) parameters
we must try and fit 13 observables:
nine charged-fermion masses and four parameters of the CKM matrix.
Even if there is,
in the complex case,
an excessive number of parameters,
the fitting may prove impossible,
due to the fact that a large number of those parameters are phases.

\section{Fermion masses and invariants}
\label{masses+invariants}

We first confine ourselves to the masses.
For brevity of notation we introduce
\be
\begin{array}{rclcrclcrcl}
\sigma_d &=& m_d^2 + m_s^2 + m_b^2, & & 
\rho_d &=& m_d^2 m_s^2 + m_s^2 m_b^2 + m_b^2 m_d^2, & &
\pi_d &=& m_d^2 m_s^2 m_b^2, \\
\sigma_\ell &=& m_e^2 + m_\mu^2 + m_\tau^2, & &
\rho_\ell &=& m_e^2 m_\mu^2 + m_\mu^2 m_\tau^2 + m_\tau^2 m_e^2, & &
\pi_\ell &=& m_e^2 m_\mu^2 m_\tau^2, \\
\sigma_u &=& m_u^2 + m_c^2 + m_t^2, & &
\rho_u &=& m_u^2 m_c^2 + m_c^2 m_t^2 + m_t^2 m_u^2, & &
\pi_u &=& m_u^2 m_c^2 m_t^2.
\end{array}
\label{invariants}
\ee
We define the matrices 
$H_a \equiv M_a M_a^\dagger$ ($a = d,\,\ell,\,u$), 
which have eigenvalue equations
\be
\det \left( m^2 \mathbbm{1} - H_a \right) = 
m^6 - \sigma_a m^4 + \rho_a m^2 - \pi_a = 0.
\ee
With the mass matrices~(\ref{m}) we obtain the relations
\ba
\sigma_d &=& s_2 + 2 a_2, \label{I1d} \\
\sigma_\ell &=& s_2 + 2 r^2 a_2, \label{I1l} \\
\sigma_u &=& p^2 s_2 + 2 q^2 a_2, \label{I1u}
\ea
where
\ba
s_2 &=& \mbox{tr} \left( S S^\ast \right), \\
a_2 &=& -\frac{1}{2}\, \mbox{tr} \left( A A^\ast \right);
\ea
also,
\ba
\rho_d &=& s_4 + a_2^2 + 2 z_4
+ 2\, \mbox{Re} \left( e^{2 i \psi} \bar z_4 \right),
\label{I2d} \\
\rho_\ell &=& s_4 + r^4 a_2^2 + 2 r^2 z_4
+ 2 r^2\, \mbox{Re} \left( e^{2 i \theta} \bar z_4 \right),
\label{I2l} \\
\rho_u &=& p^4 s_4 + q^4 a_2^2 + 2 p^2 q^2 z_4
+ 2 p^2 q^2\, \mbox{Re} \left( e^{2 i \xi} \bar z_4 \right),
\label{I2u}
\ea
where
\ba
s_4 &=& \frac{1}{2} \left[ s_2^2
- \mbox{tr} \left( S S^\ast S S^\ast \right) \right], \\
z_4 &=& s_2 a_2 + \mbox{tr} \left( S S^\ast A A^\ast \right), \\
\bar z_4 &=& - \frac{1}{2}\, \mbox{tr} \left( A S^\ast A S^\ast \right);
\ea
finally,
\ba
\pi_d &=& \left| s_3 + e^{2 i \psi} z_3 \right|^2,
\label{I3d} \\
\pi_\ell &=& \left| s_3 + r^2 e^{2 i \theta} z_3 \right|^2,
\label{I3l} \\
\pi_u &=& \left| p^3 s_3 + p q^2 e^{2 i \xi} z_3 \right|^2,
\label{I3u}
\ea
where
\ba
s_3 &=& \det S, \\
z_3 &=& \mbox{tr} \left( S A^2 \right) - \frac{1}{2}\,
\mbox{tr}\, S\, \mbox{tr} \left( A^2 \right).
\ea

\section{$b$--$\tau$ unification}
\label{inequalities}

In~\cite{bajc06} the mass matrices for the charged-fermion sector
are the same as in this paper,
but the discussion is confined to the two-generations case.
In that paper,
approximate $b$--$\tau$ unification is traced back to some inequalities
derived from the specific structure of the mass matrices.
Here we show that analogous inequalities
hold in the three-generations case.

It is convenient to use the weak basis of Eq.~(\ref{basis1}).
We remind that,
in that weak basis,
$a$,
$b$,
and $c$ are real and non-negative,
while $x$,
$y$,
and $z$ are in general complex.
One has
\ba
s_2 &=& a^2 + b^2 + c^2, \label{s2} \\
a_2 &=& \left| x \right|^2 + \left| y \right|^2 + \left| z \right|^2,
\label{a2} \\
s_4 &=& a^2 b^2 + b^2 c^2 + c^2 a^2, \\
z_4 &=& a^2 \left| x \right|^2 + b^2 \left| y \right|^2
+ c^2 \left| z \right|^2, \label{z4} \\
\bar z_4 &=& b c x^2 + c a y^2 + a b z^2, \label{barz4} \\
s_3 &=& a b c, \\
z_3 &=& a x^2 + b y^2 + c z^2.
\ea
Note that $z_3$ and $\bar z_4$ are in general complex,
while the other parameters are real.

From Eqs.~(\ref{I2u}),
(\ref{barz4}),
and~(\ref{z4}) we derive
\ba
\rho_u &\geq& p^4 s_4 + q^4 a_2^2 + 2 p^2 q^2 z_4
- 2 p^2 q^2 \left( b c \left| x \right|^2 + c a \left| y \right|^2
+ a b \left| z \right|^2 \right)
\\ &=& p^4 s_4 + q^4 a_2^2 + 2 p^2 q^2 \left[
\left( a^2 - b c \right) \left| x \right|^2
+ \left( b^2 - c a \right) \left| y \right|^2
+ \left( c^2 - a b \right) \left| z \right|^2 \right].
\label{rho1}
\ea
Without loss of generality we assume that
\be
\label{cond3}
b \geq a, \quad c \geq a.
\ee
Since $a+b+c$ is non-negative,
the inequalities~(\ref{cond3}) are equivalent to
\be
\label{cond1}
b^2 - ca \geq a^2 - bc, \quad c^2 - ab \geq a^2 - bc.
\ee
Applying the inequalities~(\ref{cond1})
to the inequality~(\ref{rho1})
and remembering Eq.~(\ref{a2}),
we obtain
\be
\rho_u \geq p^4 s_4 + q^4 a_2^2 + 2 p^2 q^2 \left( a^2 - bc \right) a_2.
\label{rho2}
\ee

We next rewrite Eqs.~(\ref{I1u}) and~(\ref{s2}) as
\be
\label{sigma1}
a_2 = \frac{1}{2 q^2} \left[ \sigma_u
- p^2 \left( a^2 + b^2 + c^2 \right) \right].
\ee
We plug this equation into inequality~(\ref{rho2})
and find after some algebra that
\be
\label{ineq}
\rho_u \geq \frac{1}{4} \left[ \sigma_u -
p^2 \left( b + c \right)^2 \right]^2 + F,
\ee
where
\be
F = \frac{p^2 a^2}{2} \left[ \sigma_u
+ p^2 \left( b + c \right)^2 - \frac{3 p^2 a^2}{2} \right].
\ee
The inequalities~(\ref{cond3}) give $b + c \geq 2 a$,
hence
\ba
F &\geq& \frac{p^2 a^2}{2} \left( \sigma_u + \frac{5 p^2 a^2}{2} \right)
\\ &\geq& \frac{p^2 a^2 \sigma_u}{2} \label{F1}
\\ &\geq& 0. \label{F}
\ea
From inequalities~(\ref{ineq}) and~(\ref{F}),
\be
\label{ineq1}
\rho_u \geq \frac{1}{4} \left[ \sigma_u -
p^2 \left( b + c \right)^2 \right]^2.
\ee
This inequality may equivalently be written
\be
\label{inequ}
\sigma_u - 2 \sqrt{\rho_u} \leq 
p^2 \left( b + c \right)^2
\leq \sigma_u + 2 \sqrt{\rho_u}.
\ee

It is obvious that,
in an exactly analogous fashion,
one may derive
\be
\begin{array}{rcccl}
\sigma_d - 2 \sqrt{\rho_d} &\leq& 
\left( b + c \right)^2
&\leq& \sigma_d + 2 \sqrt{\rho_d},
\\*[1mm]
\sigma_\ell - 2 \sqrt{\rho_\ell} &\leq& 
\left( b + c \right)^2
&\leq& \sigma_\ell + 2 \sqrt{\rho_\ell}.
\end{array}
\label{ineqs}
\ee
Inequalities~(\ref{ineqs}) should be compared with those of~\cite{bajc06}.
One reaches the same conclusion as in~\cite{bajc06}:
the intervals 
$[\sigma_d - 2 \sqrt{\rho_d},\, \sigma_d + 2 \sqrt{\rho_d}]$ and
$[\sigma_\ell - 2 \sqrt{\rho_\ell},\, \sigma_\ell + 2 \sqrt{\rho_\ell}]$
must overlap.
This overlap---at the GUT scale---implies that,
at that scale,
$m_b \simeq m_\tau$.
Notice that this conclusion was reached
without making use of the quantities $\pi_a$.

Comparing inequalities~(\ref{inequ}) and~(\ref{ineqs}),
one also finds that the parameter $p$ is approximately given by 
\be
\label{p}
p \simeq \frac{m_t}{m_b}
\ee
at the GUT scale.

Inequality~(\ref{ineq}) also delivers $F \leq \rho_u$.
Taking into account inequality~(\ref{F1}),
one has
\be
p^2 a^2 \leq \frac{2 \rho_u}{\sigma_u}.
\ee
With Eq.~(\ref{p}) in mind,
this gives,
approximately,
\be
a \lesssim \frac{\sqrt{2}\, m_c m_b}{m_t}.
\ee
Numerically,
using the values of the quark masses
in the MSSM at the GUT scale,
as given in~\cite{das}, 
one obtains for instance
$a \lesssim 3.8\, \mathrm{MeV}$ for $\tan\beta = 10$.

\section{Analytical treatment of the real case}
\label{luis}

In this section we analyze the case
of real Yukawa-coupling matrices,
i.e.\ the case of real $S$ and $A$.
In this case it is convenient to define
\ba
x_1 &=& \mbox{tr}\, S,
\\
x_2 &=& \frac{1}{2} \left[ x_1^2 - \mbox{tr} \left( S^2 \right) \right].
\ea
Then,
\ba
s_2 &=& \mbox{tr} \left( S^2 \right)
\no &=& x_1^2 - 2 x_2, \label{x2}
\\
s_4 &=& x_2^2 - 2 x_1 s_3.
\ea
With $S$ and $A$ real,
$\bar z_4$ is real,
and moreover it is not independent from $z_4$,
rather
\be
\bar z_4 - z_4 = x_2 a_2 - x_1 z_3.
\label{relation}
\ee
This allows one to write Equations~(\ref{I2d}) and~(\ref{I2l}) as
\ba
\rho_d &=& x_2^2 - 2 x_1 s_3 + a_2^2 + 2 z_4 +
2 \cos{\left( 2 \psi \right)} \bar z_4
\no &=& x_2^2 - 2 x_1 s_3 + a_2^2 - 2 x_2 a_2 + 2 x_1 z_3 +
2 \left[ 1 + \cos{\left( 2 \psi \right)} \right] \bar z_4
\no &=& \left( x_2 - a_2 \right)^2 + 2 x_1 \left( z_3 - s_3 \right)
+ 2 \left[ 1 + \cos{\left( 2 \psi \right)} \right] \bar z_4,
\label{rhod} \\
\rho_\ell &=& x_2^2 - 2 x_1 s_3 + r^4 a_2^2 + 2 r^2 z_4 +
2 r^2 \cos{\left( 2 \theta \right)} \bar z_4
\no &=& x_2^2 - 2 x_1 s_3 + r^4 a_2^2 - 2 r^2 x_2 a_2 + 2 r^2 x_1 z_3 +
2 r^2 \left[ 1 + \cos{\left( 2 \theta \right)} \right] \bar z_4
\no &=& \left( x_2 - r^2 a_2 \right)^2 + 2 x_1 \left( r^2 z_3 - s_3 \right)
+ 2 r^2 \left[ 1 + \cos{\left( 2 \theta \right)} \right] \bar z_4.
\label{rhol}
\ea
Now,
plugging Equation~(\ref{x2}) into Equations~(\ref{I1d})
and~(\ref{I1l}),
one obtains
\ba
\sigma_d &=& x_1^2 - 2 \left( x_2 - a_2 \right),
\\
\sigma_\ell &=& x_1^2 - 2 \left( x_2 - r^2 a_2 \right).
\ea
Hence,
Equations~(\ref{rhod}) and~(\ref{rhol})
may be rewritten as
\ba
\rho_d - \frac{1}{4} \left( x_1^2 - \sigma_d \right)^2
&=& 2 x_1 \left( z_3 - s_3 \right)
+ 2 \left[ 1 + \cos{\left( 2 \psi \right)} \right] \bar z_4,
\label{r1} \\
\rho_\ell - \frac{1}{4} \left( x_1^2 - \sigma_\ell \right)^2
&=& 2 x_1 \left( r^2 z_3 - s_3 \right)
+ 2 r^2 \left[ 1 + \cos{\left( 2 \theta \right)} \right] \bar z_4.
\label{r2}
\ea
In the trivial case $\cos{\left( 2 \psi \right)}
= \cos{\left( 2 \theta \right)} = -1$,
the mass matrices $M_d$ and $M_\ell$ are Hermitian
and their eigenvalues directly yield the fermion masses.
Discarding that rather trivial case from consideration,
we find that Equations~(\ref{r1}) and~(\ref{r2})
lead to
\ba
0 &=&
r^2 \left[ 1 + \cos{\left( 2 \theta \right)} \right]
\left[ \rho_d - \frac{1}{4} \left( x_1^2 - \sigma_d \right)^2
+ 2 x_1 \left( s_3 - z_3 \right) \right]
\no & &
- \left[ 1 + \cos{\left( 2 \psi \right)} \right]
\left[ \rho_\ell - \frac{1}{4} \left( x_1^2 - \sigma_\ell \right)^2
+ 2 x_1 \left( s_3 - r^2 z_3 \right) \right].
\label{master}
\ea

On the other hand,
since $s_3$ and $z_3$ are real when the matrices $S$ and $A$ are real,
Equations~(\ref{I3d}) and~(\ref{I3l}) read in that case
\be
\begin{array}{rcl}
{\displaystyle
s_3^2 + z_3^2 + 2 \cos{\left( 2 \psi \right)} s_3 z_3} &=& \pi_d,
\\*[1mm]
{\displaystyle
s_3^2 + r^4 z_3^2 + 2 r^2 \cos{\left( 2 \theta \right)} s_3 z_3} &=&
\pi_\ell.
\end{array}
\label{system}
\ee
Defining
\ba
f_1 &=& 1 - r^4,
\\
f_2 &=& \cos{\left( 2 \psi \right)} - r^2 \cos{\left( 2 \theta \right)},
\\
f_3 &=& r^4 \cos{\left( 2 \psi \right)} - r^2 \cos{\left( 2 \theta \right)},
\\
f_4 &=& \pi_d - \pi_\ell,
\\
f_5 &=& r^4 \pi_d - \pi_\ell,
\\
f_6 &=& r^2 \cos{\left( 2 \theta \right)} \pi_d
- \cos{\left( 2 \psi \right)} \pi_\ell,
\ea
the system of equations~(\ref{system}) has solutions given by
\ba
s_3^2 &=& \frac{- f_1 f_5 - 2 f_3 f_6 \pm 2 f_3 \sqrt{f_6^2 - f_4 f_5}}
{f_1^2 + 4 f_2 f_3},
\no
z_3^2 &=& \frac{f_1 f_4 - 2 f_2 f_6 \mp 2 f_2 \sqrt{f_6^2 - f_4 f_5}}
{f_1^2 + 4 f_2 f_3},
\label{s3z3} \\
s_3 z_3 &=& \frac{f_2 f_5 + f_3 f_4 \pm f_1 \sqrt{f_6^2 - f_4 f_5}}
{f_1^2 + 4 f_2 f_3}.
\nonumber
\ea

We use as input the three charged-lepton masses,
the three down-type-quark masses,
and also $r$,
$\cos{\left( 2 \theta \right)}$,
and $\cos{\left( 2 \psi \right)}$.
Equations~(\ref{s3z3}) allow us to compute $s_3$ and $z_3$
from that input.
Inserting those values of $s_3$ and $z_3$
in Equation~(\ref{master}),
we obtain a quartic equation for $x_1$,
which may be analytically solved.
The quantities $s_2$ and $a_2$ are then computed as
\ba
s_2 &=& \frac{x_1^2}{2} + \frac{r^2 \sigma_d - \sigma_\ell}
{2 \left( 1 - r^2 \right)},
\\
a_2 &=& \frac{\sigma_d - \sigma_\ell}{2 \left( 1 - r^2 \right)}.
\ea
Finally,
$\bar z_4$ is computed from either Equation~(\ref{r1})
or Equation~(\ref{r2}),
and $z_4$ is obtained from Equation~(\ref{relation}).
All the invariants pertaining to the matrices $S$ and $A$
are thus analytically computed from the input.

One must,
yet,
take into account the fact that
those invariants must satisfy several inequalities.
In the weak basis~(\ref{basis1}),
\ba
x_1 &=& a+b+c, \\
x_2 &=& a b + b c + c a, \\
s_3 &=& a b c.
\ea
The numbers $a$,
$b$,
and $c$ are real,
and they may be negative,
see footnote~\ref{thefoot}.
The quantity
\ba
\Delta &\equiv&
x_1^2 x_2^2 + 18 x_1 x_2 s_3 - 4 x_2^3 - 4 x_1^3 s_3 - 27 s_3^2
\\ &=&
\left[ \left( a - b \right)
\left( b - c \right) \left( c - a \right) \right]^2
\ea
must therefore be non-negative.
Further non-negative quantities may be conveniently derived
by using the weak basis~(\ref{basis2}) and deriving,
in that basis,
the values of $a$,
$b$,
$c$,
$d^2$,
and $f^2$ from the invariants.
From the condition that $f^2$ must be non-negative one obtains
\ba
\Sigma &\equiv& a_2 z_4 - z_3^2
\\ &\geq& 0.
\ea
From the condition that $d^2$ must be non-negative one obtains
\ba
\Psi &\equiv& - z_4^3 + z_4^2 \left( 2 x_1 z_3 - x_2 a_2 \right)
+ z_4 \left[ a_2 z_3 \left( 3 s_3 + x_1 x_2 \right)
-x_1 s_3 a_2^2 - \left( x_1^2 + x_2 \right) z_3^2 \right]
\no & &
+ z_3^3 \left( x_1 x_2 - s_3 \right)
- a_2 z_3^2 \left( x_2^2 + x_1 s_3 \right)
+ 2 x_2 s_3 a_2^2 z_3
- s_3^2 a_2^3
\\ &\geq& 0.
\ea
The conditions that $\Delta$,
$\Sigma$,
and $\Psi$ be non-negative constitute a severe constraint
on the inputted values of the charged-fermion masses and of $r$,
$\theta$,
$\psi$.

After having computed the invariants,
one may further input the three up-type-quark masses
and therefrom derive the values of $p^2$,
$q^2$,
and $\cos{\left( 2 \xi \right)}$.
In practice,
this involves solving a cubic equation,
and thereafter imposing the constraints $p^2 \geq 0$,
$q^2 \geq 0$, 
and $\left| \cos{\left( 2 \xi \right)} \right| \leq 1$.
This obviously translates
into constraints on the inputted up-type-quark masses.

In the way delineated in this section,
one may analytically solve the case of real $S$ and $A$ matrices,
by inputting the charged-fermion masses
and therefrom deriving $S$ and $A$,
without having to have recourse to fits.
In practice,
however,
doing things the other way round---trying to fit
the charged-fermion masses numerically
from some inputted values of $S$,
$A$,
and the other parameters---proves more effective.
We turn to that procedure in the next section.

\section{The fitting procedure}
\label{fitting}

In order to check whether the mass matrices~(\ref{m})
allow to reproduce the masses and CKM mixing angles
at the GUT scale,
we use a $\chi^2$ analysis,
as was previously applied for instance in~\cite{raby,schwetz}.
As for the masses, 
the $\chi^2$-function is given by 
\be
\chi^2_\mathrm{masses} = \chi^2_d + \chi^2_\ell + \chi^2_u,
\ee
where
\be
\chi^2_d = \sum_{i = d,s,b} 
\left( \frac{m_i(x) - \bar m_i}{\delta m_i} \right)^2,
\ee
and analogously for $\chi^2_{\ell,u}$.
The masses at the GUT scale are $\bar m_i \pm \delta m_i$,
whereas the $m_i(x)$ are the masses calculated from Eqs.~(\ref{m})
as functions of the parameter set
$x = \left\{ S, A, p, q,r, \psi, \theta, \xi \right\}$
(see Section~\ref{masses} for the distinction between the ``real''
and the ``complex'' cases).
The total $\chi^2$-function is the sum
\be
\chi^2_\mathrm{total} = \chi^2_\mathrm{masses} + \chi^2_\mathrm{CKM},
\ee
with
\be
\chi^2_\mathrm{CKM} = \sum_{i = 12,13,23} 
\left( \frac{ \sin \theta_i(x) - \sin \bar\theta_i}{\delta \sin \theta_i} 
\right)^2.
\ee
We take the masses $\bar m_i$ at the GUT scale,
and their errors $\delta m_i$,
from Table~II of~\cite{das};
those masses refer to the MSSM with $\tan \beta = 10$
and a GUT scale of $2 \times 10^{16}\, \mathrm{GeV}$
and have been obtained through the renormalization-group evolution
of the masses given in~\cite{fusaoka} at the $Z^0$-mass scale.
As for sines of the CKM angles,
$\sin \bar\theta_i \pm \delta \sin \theta_i$, 
we use Table~1 in~\cite{malinsky}.
We do not take into account the CKM phase in our fitting procedure;
this omission will be justified later.

In order to get a better understanding of our mass matrices,
we perform separate minimizations of $\chi^2_\mathrm{masses}$
and of $\chi^2_\mathrm{total}$. 
We also test the ``real'' versus the ``complex'' case.

For the numerical multi-dimensional minimization
of the $\chi^2$-functions we employ
the downhill simplex method~\cite{downhill}.
Because the problem is highly non-linear,
we expect the existence of many local minima.\footnote{The concept
``local minimum'' is not understood in a strict mathematical sense,
rather it refers to a point where the minimization 
algorithm successfully stops.} 
We start with randomly generated initial simplices. 
At the points where the numerical algorithm stops,
we iterate the procedure with random perturbations
in order to find a lower $\chi^2$.
In this way we can be fairly certain
about the distribution of the local minima
and about the position of the global minimum.

In the description of the fits,
the concept of ``pull'' with respect to an observable $O$ is useful.
The pull of $O$ is defined as
\be
\mbox{pull} \left( O \right) = \frac{O \left( \hat x \right)
- \bar O}{\delta O}, 
\ee
where the experimental value of the observable is $\bar O \pm \delta O$,
while $O \left( x \right)$ is the theoretical prediction of $O$,
given as a function of the parameter set $x$;
$\hat x$ is the parameter set at a local minimum of $\chi^2$.
Thus, 
\be
\min \chi^2 \left( x \right)
\equiv \chi^2 \left( \hat x \right)
= \sum_O \left[ \mbox{pull} \left( O \right) \right]^2.
\ee

\section{Results}
\label{results}

We have performed all fits and  tests of our scenario
separately for real and complex
coupling matrices $S$ and $A$---see Section~\ref{masses}. 
It turns out that there are no significant numerical differences
between the two cases.
The extra two phases in the ``complex'' case
are unable to significantly improve our fits.
Therefore,
for simplicity in the following we confine ourselves to the ``real'' case.

\begin{figure}[t]
\begin{center}
\epsfig{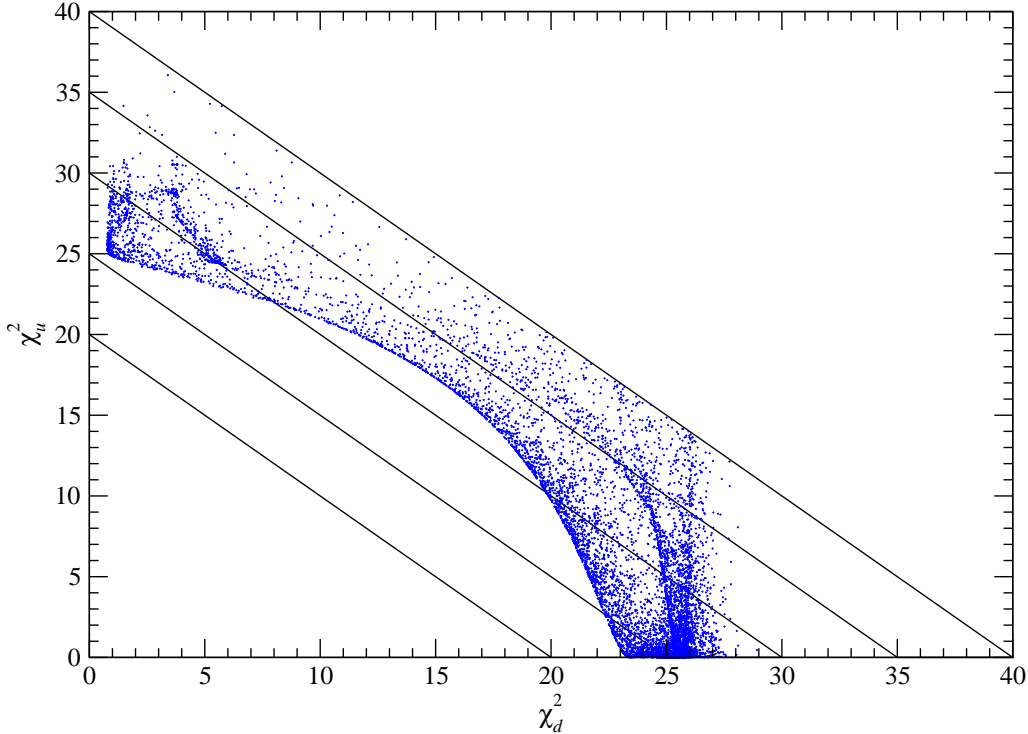}
\end{center}
\caption{The distribution of local minima of
$\chi^2_\mathrm{masses} \approx \chi^2_d + \chi^2_u$
in the $\chi^2_d$--$\chi^2_u$ plane
(we force $\chi^2_\ell$ to be always negligibly small).
The straight lines refer to constant values of $\chi^2_\mathrm{masses}$.
This figure refers to the ``real'' case.
\label{nomix}}
\end{figure}

\pagebreak
\paragraph{Fits of the masses alone:} 
Firstly we omit the CKM angles and test whether,
with the mass matrices~(\ref{m}),
we are able to fit the charged-fermion mass values at the SUSY GUT scale
given in~\cite{das}.
In Fig.~\ref{nomix} we show the distribution,
in the $\chi^2_d$--$\chi^2_u$ plane,
of the local minima of $\chi^2_\mathrm{masses}$
for which $\min \chi^2_\mathrm{masses} \leq 40$.
Though the density of points in that figure
depends sensitively on the number of random perturbations 
and on the number of restarts of the downhill simplex procedure,
the overall picture is clear.
The absolute minimum of $\chi^2_\mathrm{masses}$
is located at $\chi^2_u \simeq 0,\ \chi^2_d \simeq 23.3$; 
the corresponding fit masses, 
and the pulls, are given in Table~\ref{without}.
\begin{table}
\begin{center}
\renewcommand{\arraystretch}{1.2}
\begin{tabular}{|c|c||c|c||c|c|} \hline
\multicolumn{2}{|c||}{} & 
\multicolumn{2}{c||}{$\chi^2_\mathrm{masses} = 23.3$} & 
\multicolumn{2}{c|}{$\chi^2_\mathrm{masses} = 25.8$} \\ \hline
 & $\bar m$ & 
$m \left( \hat x \right)$ & $\mbox{pull} \left( m \right)$ & 
$m \left( \hat x \right)$ & $\mbox{pull} \left( m \right)$ \\ \hline
$m_e$    & 0.3585  & 0.3585 & $6 \times 10^{-3}$ 
                   & 0.3585 & $6 \times 10^{-4}$ \\
$m_\mu$  & 75.67   & 75.67  & $-4 \times 10^{-4}$ 
                   & 75.67  & $-9 \times 10^{-4}$ \\
$m_\tau$ & 1292.2  & 1292.2 & $-4 \times 10^{-3}$ 
                   & 1292.2 & $-4 \times 10^{-3}$ \\
$m_d$    & 1.504   & 0.4112 & $-4.74$  
                   & 1.430  & $-0.321$ \\
$m_s$    & 29.95   & 29.54  & $-0.090$ 
                   & 29.35 & $-0.132$ \\
$m_b$    & 1063.6  & 1187.0 & 0.873  
                   & 1188.2 & 0.882 \\
$m_u$    & 0.7238  & 0.7249 & $8 \times 10^{-3}$ 
                   & 0.7321 & 0.061 \\
$m_c$    & 210.33  & 212.47 & 0.113 
                   & 214.66 & 0.228 \\
$m_t$    & 82433   & 78466  & $-0.269$ 
                   &  8778  & $-4.99$ \\
\hline
\end{tabular}
\end{center}
\caption{Results of the fit for the ``real'' case without the CKM angles.
The values of the masses in the second column,
and the corresponding errors $\delta m$ for the calculation of the pulls,
have been taken from~\cite{das}.
The third column gives our best fit
and the fourth column displays the corresponding pulls.
The fifth and sixth columns refer to best fit
in the region of small $\chi^2_d$,
i.e.\ the region in Fig.~\ref{nomix} with $\chi^2_d \simeq 1$
and $\chi^2_u \simeq 25$.
All the masses are in units of MeV.
\label{without}}
\end{table}
For comparison,
we also show in Table~\ref{without}
the central mass values of~\cite{das}.
Looking at the pulls,
we see that this mass fit fails only in the mass of the down quark;
that particular pull is responsible
for almost the complete $\chi^2_\mathrm{masses} = 23.3$.
A glance at Fig.~\ref{nomix} also reveals
that there are local minima with $\chi^2_u \simeq 25$
and $\chi^2_d \simeq 1$;
those minima,
the best of which is also displayed in Table~\ref{without},
give rather good fits for \emph{all} the down-type-quark masses,
but fail severely in fitting the top-quark mass:
the fit value is about one order of magnitude,
or five $\sigma$,
smaller than the experimental value---see Table~\ref{without}.

Thus,
with our scenario we cannot even fit all the charged-fermion masses.
However,
as stressed in the introduction,
our scenario is extreme in that it allows only for the Yukawa couplings
of one $\mathbf{10} \oplus \mathbf{120}$ scalar representation.
If we allow for small perturbations of the mass matrices,
there are several ways out:
there could be contributions
from Yukawa couplings of one $\overline{\mathbf{126}}$~\cite{aulakh06},
several $\mathbf{10}$ and/or $\mathbf{120}$ of scalars, 
radiative corrections,
or non-renormalizable terms.
Consequently,
the absolute minimum of $\chi^2_\mathrm{masses}$ can be considered acceptable,
since it fails only for $m_d$,
which is small anyway.
On the other hand,
our philosophy of small perturbations
forces us to discard the local minimum
where the fit value of $m_t$ is one order of magnitude too small.

\paragraph{Fits with masses and CKM angles:}

\begin{figure}[t]
\begin{center}
\epsfig{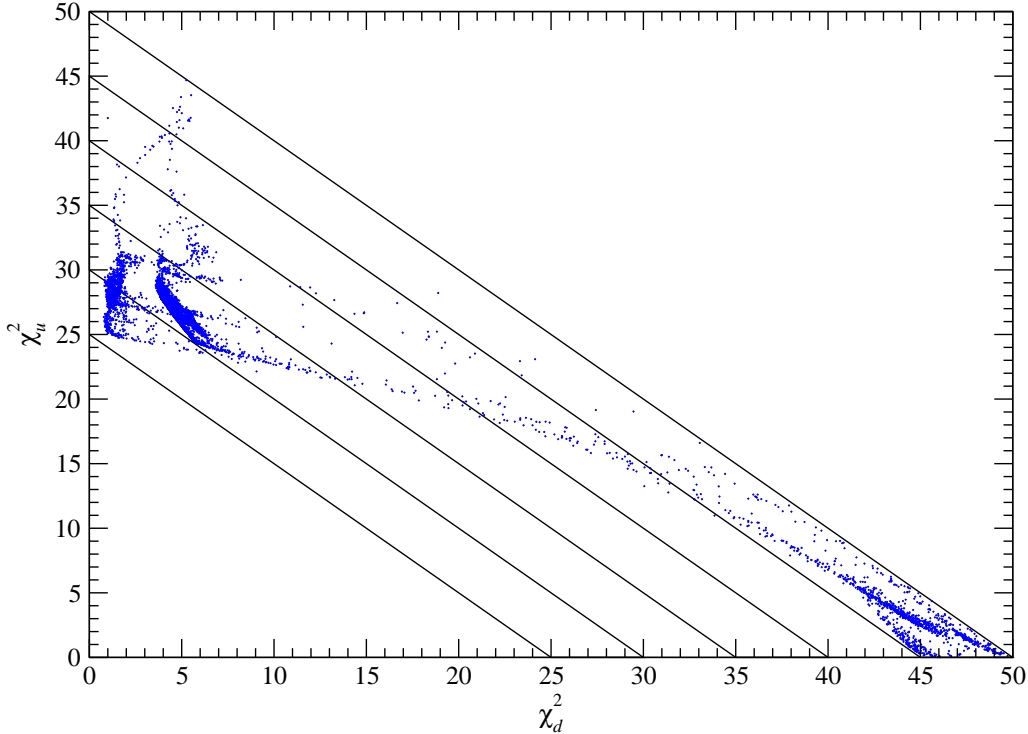}
\end{center}
\caption{The distribution of local minima of $\chi^2_\mathrm{total}$ 
(in the ``real'' case)
in the $\chi^2_d$--$\chi^2_u$ plane.
The straight lines refer to constant
$\chi^2_d + \chi^2_u \simeq \chi^2_\mathrm{total}$.
\label{mix}}
\end{figure}

Figure~\ref{mix} shows the distribution
in the $\chi^2_d$--$\chi^2_u$ plane,
for the ``real'' case,
of the local minima of $\chi^2_\mathrm{total}$
which have $\min \chi^2_\mathrm{total} \leq 50$.
We see that the gross feature---the lower left corner is devoid of local
minima---is the same as in the fit without CKM angles.
The previous local minimum at $\chi^2_u \simeq 25$
and $\chi^2_d \simeq 1$ is now the absolute minimum.
That absolute minimum is given in detail in Table~\ref{with}.
\begin{table}[t]
\begin{center}
\renewcommand{\arraystretch}{1.2}
\begin{tabular}{|c|c||c|c||c|c|} \hline
\multicolumn{2}{|c||}{} & 
\multicolumn{2}{c||}{$\chi^2_\mathrm{total} = 26.9$} &
\multicolumn{2}{c||}{$\chi^2_\mathrm{total} = 45.4$}
\\ \hline
 & $\bar m$ & 
$m \left( \hat x \right)$ & $\mbox{pull} \left( m \right)$ &
$m \left( \hat x \right)$ & $\mbox{pull} \left( m \right)$
\\ \hline
$m_e$    & 0.3585  & 0.3585 & $-1 \times 10^{-3}$ 
                   & 0.3585 &  $7 \times 10^{-3}$ \\
$m_\mu$  & 75.67   & 75.67  &  $2 \times 10^{-3}$ 
                   & 75.67  &  $4 \times 10^{-3}$ \\
$m_\tau$ & 1292.2  & 1292.2 & $-4 \times 10^{-3}$ 
                   & 1292.2 & $-3 \times 10^{-3}$ \\
$m_d$    & 1.504   & 1.563  & 0.141 
                   & 1.044  & $-1.993$            \\
$m_s$    & 29.95   & 28.24  & $-0.376$ 
                   & 1.36   & $-6.29$             \\
$m_b$    & 1063.6  & 1191.0 & 0.903 
                   & 1225.4 & 1.145               \\
$m_u$    & 0.7238  & 0.7243 & $3 \times 10^{-3}$  
                   & 0.7279 & 0.030               \\
$m_c$    & 210.33  & 215.35 & 0.264 
                   & 216.83 & 0.342               \\
$m_t$    & 82433   & 8179   & $-5.03$ 
                   & 74145  & $-0.56$                \\ \hline
 & $\sin \bar\theta$ & $\sin \theta \left( \hat x \right)$ &
$\mbox{pull} \left( \sin \theta \right)$ &
$\sin \theta \left( \hat x \right)$ &
$\mbox{pull} \left( \sin \theta \right)$ 
\\ \hline
$\sin \theta_{12}$ & 0.2243 & 0.2242 & $-0.047$ 
& 0.2243 & $2 \times 10^{-4}$ \\
$\sin \theta_{23}$ & 0.0351 & 0.0348 & $-0.208$ 
& 0.0352 & 0.093 \\
$\sin \theta_{13}$ & 0.0032 & 0.0036 & 0.740 
& 0.0034 & 0.318 \\
\hline
\end{tabular}
\end{center}
\caption{
Results of the fit for the ``real'' case including the CKM angles.
The best fit is described by columns three and four: this point lies
at $\chi^2_d \simeq 1$, $\chi^2_u \simeq 25$. Columns five and six
refer to the best fit in the region of small $\chi^2_u$.
\label{with}}
\end{table}
We see that the fit of $m_t$ is as unacceptably bad as before,
but the CKM angles are well reproduced.

Moving to the zone of $\chi^2_u \lesssim 1$ in Fig.~\ref{mix},
where the top-quark mass is well reproduced,
we find the following characteristic features:
\begin{itemize}
\item
In that zone the best fit has $\chi^2_\mathrm{total} \simeq 45.4$.
The corresponding fit values and pulls are shown in
columns
five and six of Table~\ref{with}, respectively.
\item
Varying $\chi^2_\mathrm{total}$ between 45.4 and 49, we find that
the pull of $m_d$ changes roughly from $-2$ to $-3$.
\item
For that range of $\chi^2_\mathrm{total}$,
the pull of $m_s$ remains close to $-6$,
i.e.\ the fit value of $m_s$
is one order of magnitude lower than the experimental value---indeed,
$m_s$ turns out hardly larger than $m_d$!
This is the main reason why $\chi^2_\mathrm{total}$ is so bad
in the region of low $\chi^2_u$.
\item
The pull of $m_b$ is always about $+1$.
\end{itemize}

Since we are unable to reproduce well all the quark masses,
we cannot expect to obtain a realistic CKM phase,
and we have not included it in our fit.

In view of our philosophy,
we have also tried a fit of the masses and of the CKM mixing angles
while allowing for artificially large errors in the light-fermion masses. 
Taking for instance $\delta m_i = 5\, \mathrm{MeV}$ for $i = e,\, d,\, u$, 
we are able to achieve $\chi^2_\mathrm{total} \simeq 25.4$.
This is not really an improvement
when compared to the best fit in columns 3 and 4 of Table~\ref{with}. 
However,
the characteristics of this fit are different from those of that best fit:
the pulls of $m_b$ and of $m_t$ are approximately $+1$ and $-1$,
respectively,
whereas the fit value of $m_s$ is $4.65\, \sigma$ too low.
Thus,
the fit with drastically increased errors in the light-fermion masses
rather resembles the fit of columns 5 and 6 of Table~\ref{with}.

\paragraph{A numerical test of $b$--$\tau$ unification:}

\begin{figure}
\begin{center}
\epsfig{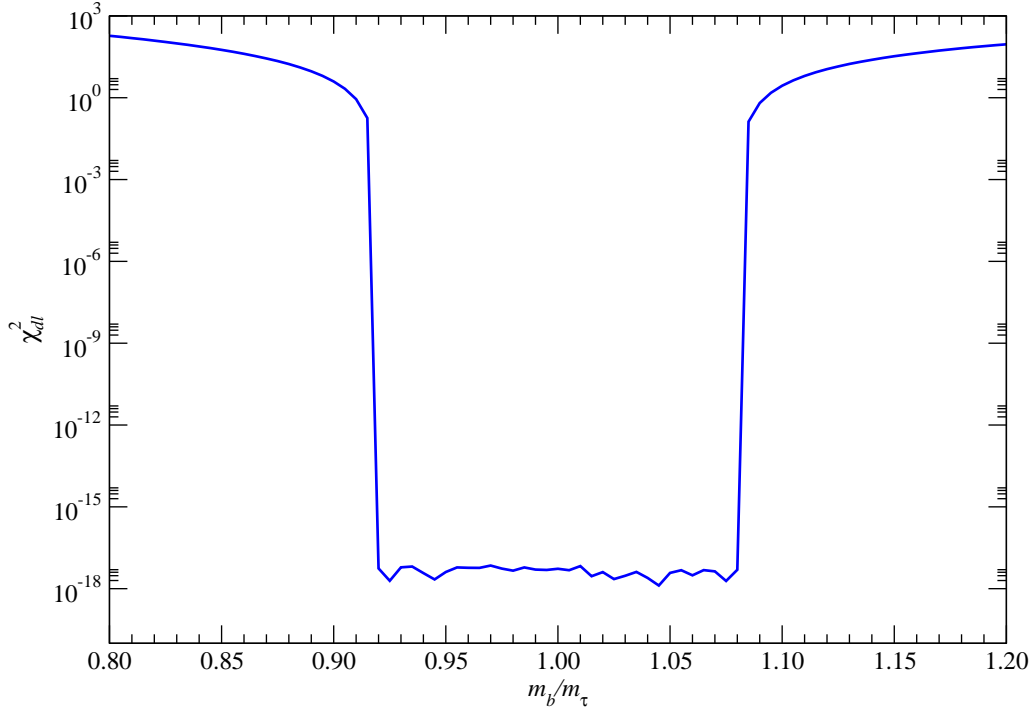}
\end{center}
\caption{The minimum of the $\chi^2$-function of Eq.~(\ref{chi2(mb)}) as a
  function of $m_b$.
\label{mb-chisqr}}
\end{figure}

In Section~\ref{inequalities}
we have traced $b$--$\tau$ unification
to some inequalities involving the charged-lepton
and the down-type-quark masses;
those inequalities are conditions on the masses
necessary for Eqs.~(\ref{I1d}),
(\ref{I1l}),
(\ref{I2d}),
and~(\ref{I2l}) to have a solution.
Neglecting the masses of the first generation,
i.e.\ setting $m_d \simeq m_e \simeq 0$,
those conditions are reformulated as~\cite{bajc06}
\be
\label{mb-mtau}
1 - \frac{m_\mu + m_s}{m_\tau}
\lesssim \frac{m_b}{m_\tau} \lesssim
1 + \frac{m_\mu + m_s}{m_\tau}.
\ee
Using the values $\bar m_i$ for the masses in~\cite{das},
this reads
\be
\label{bounds}
0.92 \lesssim \frac{m_b}{m_\tau} \lesssim 1.08.
\ee
We have performed a $\chi^2$ analysis
to check the inequalities~(\ref{mb-mtau}).
For this purpose, 
we consider
\be
\label{chi2ld}
\chi^2_{d\ell} \left( x,m_b \right) \equiv 
\sum_{i = e, \mu, \tau, d, s} 
\left( \frac{m_i \left( x \right) - \bar m_i}{\delta m_i} \right)^2 + 
\left( \frac{m_b \left( x \right) - m_b}{0.01\, m_b} \right)^2.
\ee
This $\chi^2$-function is identical
with $\chi^2_d + \chi^2_\ell$
apart from the term corresponding to the bottom-quark mass,
wherein we leave $m_b$ free
and assign to it a very small error bar of 1\%.
Then we define a minimal $\chi^2$ as a function of $m_b$:
\be
\label{chi2(mb)}
\chi^2_{d\ell} \left( m_b \right)
\equiv \min_x \chi^2_{d\ell} \left( x,m_b \right).
\ee
This function allows one to test the down-type-quark
and charged-lepton mass fits
with respect to variations of $m_b$~\cite{schwetz}. 
In Fig.~\ref{mb-chisqr} we have plotted 
$\chi^2_{d\ell} \left( m_b \right)$
against $m_b/m_\tau$;
we have used the mean value $m_\tau = 1292.2\, \mathrm{MeV}$
given in~\cite{das}.
We see that exactly in the range of Eq.~(\ref{bounds})
the minimum of $\chi^2_{d\ell} \left( m_b \right)$ is,
for all practical purposes,
zero.
This confirms our analytic derivation of $b$--$\tau$ unification. 

We can also test Eq.~(\ref{p}) against our numerics.
We find that that equation is reproduced
fairly well whenever the fit of $m_t$ is good.

\section{Summary} \label{concl}

In this paper we have investigated a SUSY $SO(10)$ scenario
in which the charged-fermion masses are generated exclusively 
by the renormalizable Yukawa couplings of the fermions
to one representation $\mathbf{10} \oplus \mathbf{120}$ of scalars.
We have studied the three-generations case,
confirming the $b$--$\tau$ unification
which had previously been proved for two generations~\cite{bajc06}. 
However,
our tests of this scenario
against the charged-fermion masses
and against the CKM mixing angles at the GUT scale
show that it is not satisfactory:
the fit value of $m_t$ comes out much too low for the best fit;
allowing for a larger $\chi^2_\mathrm{total}$, 
we are able to obtain a good fit of $m_t$,
at the price of $m_s$ turning out one order of magnitude too low
and of $m_d$ also being too small.
We thus find that the scenario
investigated here is too restrictive:
an additional mechanism
for charged-fermion mass generation is required.

\paragraph{Acknowledgements:}
We thank Thomas Schwetz
for advice and discussions on the fitting procedure.
The work of L.L.\ has been supported
by the Portuguese \textit{Funda\c c\~ao para a Ci\^encia e a Tecnologia}
through the project U777-Plurianual.

\end{document}